\documentclass[letter]{aa}

\usepackage{graphicx}
\usepackage{txfonts}
\usepackage{natbib}    
\bibpunct{(}{)}{;}{a}{}{,}   

\newcommand{\microns}{\ensuremath{\mu\mathrm{m}}}
\newcommand{\Tstar}{\ensuremath{T_\star}}
\newcommand{\Tlayer}{\ensuremath{T_l}}
\newcommand{\Dstar}{\ensuremath{\Phi_\star}}
\newcommand{\Dlayer}{\ensuremath{\Phi_l}}
\newcommand{\Olayer}{\ensuremath{\tau_l}}
\newcommand{\mira}{\texttt{MIRA}}

\newcommand{\base}{\ensuremath{\mathrm{B{}}}}
\newcommand{\wave}{\ensuremath{\lambda{}}}

\newcommand{\vis}{\ensuremath{\mathrm{V}}}

\newcommand{\blambda}{\ensuremath{\base{}_\wave{}}}

\newcommand{\TLep}{\object{T~Lep}}

\begin{document}

\title{Pre-maximum spectro-imaging of the Mira star T~Lep with AMBER/VLTI\thanks{Based on observations collected at the VLTI, ESO-Paranal, Chile.}}

\author{J.-B.~Le~Bouquin\inst{1} \and S.~Lacour\inst{2} \and S.~Renard\inst{2} \and E.~Thi\'ebaut\inst{3} \and A.~Merand\inst{1} \and T. Verhoelst\inst{4}}

\institute{European Southern Observatory, Casilla 19001, Santiago 19, Chile \and
LAOG--UMR 5571, CNRS and Universit\'e Joseph Fourier, BP 53, F-38041 Grenoble, France \and
AIRI/Observatoire de Lyon, France and Jean-Marie Mariotti Center, France \and
Instituut voor Sterrenkunde, KULeuven, Celestijnenlaan 200D B-3001 Leuven, Belgium}

\offprints{J.B.~Le~Bouquin \\ \email{jlebouqu@eso.org}}
  
\date{Received 24/12/2008 ; Accepted  04/02/2009}

\abstract{ Diffuse envelopes around Mira variables are among the most
  important sources influencing the chemical evolution of galaxies. However
  they represent an observational challenge because of their complex
  spectral features and their rapid temporal variability.}
{ We aimed at constraining the exact brightness distribution of the
  Mira star T~Lep with a model-independent analysis.}
{ We obtained single-epoch interferometric observations
  with a dataset continuous in the spectral domain
  ($\wave=1.5-2.4\,\microns{}$) and in the spatial domain
  (interferometric baselines ranging from $11$ to $96\,$m). We performed
  a model independent image reconstruction for each spectral bin using
  the MIRA software. We completed the analysis by modeling the data
  with a simple star+layer model inspired from the images.}
{ Reconstructed images confirm the general picture of a central star
  partially obscured by the surrounding molecular shell of changing
  opacity.  At $1.7\,\microns$, the shell becomes optically thin,
  with corresponding emission appearing as a ring
  circling the star. This is the first
  direct evidence of the spherical morphology of the molecular shell.
  Model fitting confirmed a spherical layer of constant size and changing
  opacity over the wavelengths.
  Rough modeling points to a continuum
  opacity within the shell, in addition to the CO and
  H$_2$O features. Accordingly, it appeared impossible to model 
  the data by a photosphere alone in any of the spectral bins. }
{}


\maketitle


\section{Introduction}

Among the Asymptotic Giant Branch members, Mira stars are low-mass
($1M_\odot$), large-amplitude ($\Delta{}V\approx9$), long-period
variables ($\approx{1}$yr), evolving toward the planetary nebula and
white dwarf phases. Their important mass-loss rate, of the order of
$10^{-6}M_\odot$/year, significantly affects their evolution and is
one of the most important sources for the chemical enrichment of the
interstellar medium. A better understanding of these late stages of
stellar evolution may help shed light on this important player in the
chemical evolution of galaxies.
Due to their large diameters and high luminosities, Mira variables are
a favorite target for observations at high angular resolution in the optical and
near-infrared (NIR) wavelengths. The wealth of aperture-masking and
long baseline interferometric information has strongly advanced the
study of molecule and dust formation
\citep{Perrin-1999may,Thompson-2002sep,Woodruff-2004jul,Ohnaka-2004sep,Ireland-2006apr,Ragland-2006nov}. The
global picture is an onion-like structure with molecular layers and
dust shells surrounding the photosphere, the latter animated by
various activities (pulsations, convection, wave-shocks) crucial to
explain the mass-loss process and the structural evolution.

To fully characterize the structure of a Mira star's atmosphere, one
would need the complete intensity map at all wavelengths and all
pulsation phases, recorded over a large number of pulsation
cycles. The majority of recent studies concentrated on the spectral
and/or the phase dependence. They overcame the issue of the exact
brightness distribution by using either integrated quantities
\citep[such as
  \emph{diameter},][]{Millan-Gabet-2005feb,Woodruff-2008jan}, or the
direct comparison of a few interferometric measurements with geometric
\citep{Mennesson-2002nov,Perrin-2004oct} or dynamic models of
pulsating atmospheres
\citep{Ohnaka-2006feb,Wittkowski-2007jul,Wittkowski-2008feb}. 
To study the brightness distribution, the difficulty lies in
collecting a sufficient amount of spatial information in a period of
time shorter than the typical life-time of the expected structures,
and within spectral bins small enough to spectrally resolve the
molecular bands (for which completely different intensity maps are
expected).

In this work, we present a multi-wavelength, spatially resolved
observation of the single star \TLep{}. This Mira variable
has a spectral type of M6e-M9e and V-band magnitude varying from
$7.3$ to $14.3$ \citep{Samus-2004nov}. For the ephemeris, we use a
Modified Julian Day of last maximum brightness $T_0=54446$ days and a
pulsation period of $380$ days (based on the last cycles observed by
AAVSO), instead of the $368$ days from \citet{Whitelock-2000dec}. The
mass-loss has been estimated to be
$7.3\times10^{-7}M_\odot$/year by \citet{Loup-1993jun}.
The most prominent spectral features expected in the
NIR comes from water vapor and also CO.
The shapes and widths of these molecular bands depend
on the stellar phase.

We present the dataset in Sec.~\ref{sec:observations}, with a special
emphasis on the observations and data reduction. Image reconstruction
is presented and discussed in Sec.~\ref{sec:image}.
We model the data in Sec.~\ref{sec:profile} with a simple star+layer
model inspired by \citet{Perrin-2004oct}.

\section{Observations and data reduction}
\label{sec:observations}

 Data were collected at the Very Large Telescope Interferometer
 \citep[VLTI,][]{Haguenauer-2008spie} with the spectrograph AMBER
 \citep{Petrov-2007mar} covering simultaneously the J-, H- and K-bands
 with a spectral resolution of $R\approx$$35$. Even though J-band fringes
 have been properly recorded in several observations, we decided to
 discard them from the analysis of this paper since the data quality
 is significantly worse than for longer wavelengths. The majority of
 the observations have been obtained by the FINITO fringe-tracker 
 \citep{LeBouquin-2008spie_b}. When the conditions were
 unstable, we used FINITO in group-tracking mode (instead of
 phase-tracking). This mode reduces the sensitivity to unstable
 atmospheres at the cost of a clearly reduced instrumental
 contrast. We used 4
 configurations of 3 Auxiliary Telescopes (ATs) each. Data for the
 A0-D0-H0, D0-H0-G1 and E0-G0-H0 configurations were obtained
 within a few days of each other while data with G2-G0-K0, extracted from the archive,
 was obtained about 1 month before as shown in
 Table~\ref{tab:obs_log}.

\begin{figure}
    \centering 
    \includegraphics[scale=0.6]{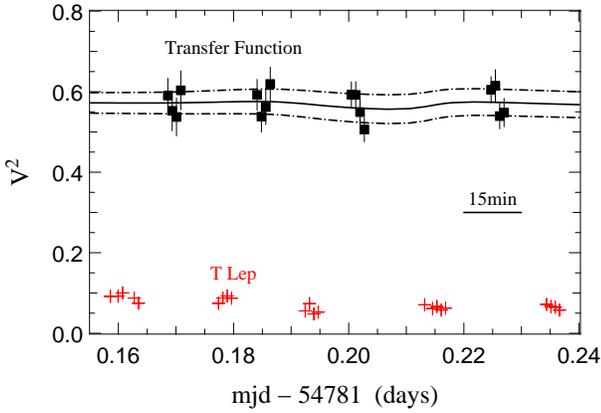}
    \caption{Example of observation sequence, on the baseline G0-H0 and for the spectral bin $1.59\,\microns$. The interpolated TF is represented.}
    \label{fig:calibrate}
\end{figure}

\begin{table}
  \centering
  \caption{Observation log.}
  \begin{tabular}[c]{cccccccc} \hline \hline \vspace{-0.29cm}\\
    Date      & \TLep{} phase & Baselines     & Spectral range 
    \vspace{0.03cm}\\ \hline \vspace{-0.30cm}\\
    2008-10-10 & $0.80$ & G2-G0-K0 & $1.5087-2.3191\,\microns$\\
    2008-11-04 & $0.86$ & D0-H0-G1 & $1.4287-2.4413\,\microns$\\
    2008-11-06 & $0.87$ & A0-D0-H0 & $1.4287-2.4238\,\microns$\\
    2008-11-10 & $0.88$ & E0-G0-H0 & $1.4287-2.4413\,\microns$\\\hline
  \end{tabular}
  \flushleft
  \label{tab:obs_log}
\end{table}

Concerning the wavelength tables, we performed a rough absolute
calibration by cross-correlating our observed spectra with an
atmospheric model. We found a systematic offset of
$0.08\pm{}0.02\,\microns{}$ with respect to the initial AMBER table.

\begin{figure}
    \centering 
    \includegraphics[scale=0.6]{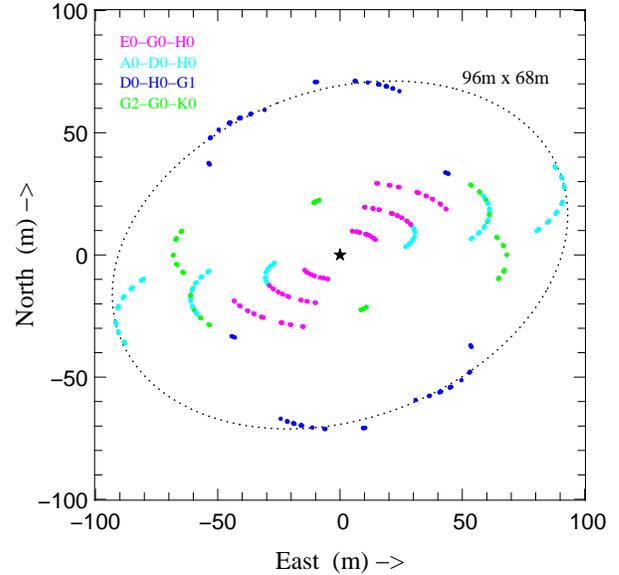}
    \caption{Projected baselines (UV-plane) in meters of the complete AMBER dataset on \TLep{}.}
    \label{fig:uv_plane}
\end{figure}

\begin{figure*}
    \centering \vspace{-0.2cm}
    \includegraphics[angle=-90,scale=0.63]{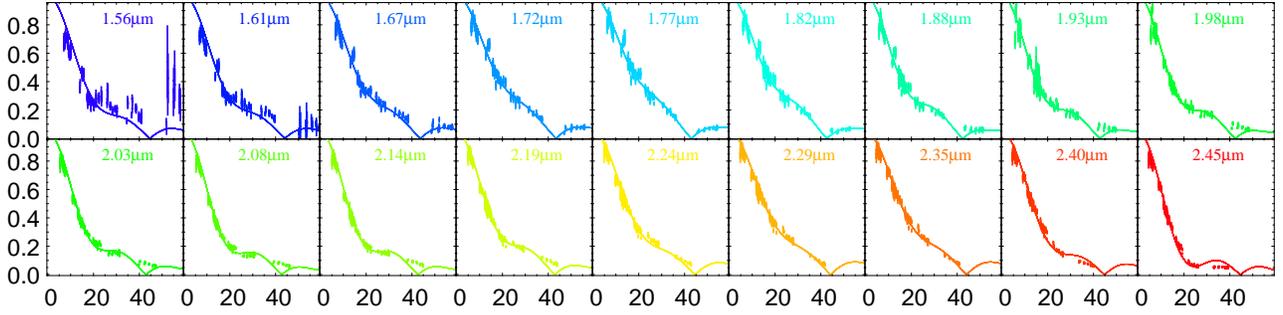}
    \caption{Visibility profiles $\vis(\blambda)$ from AMBER/VLTI,
      compared to the star+layer model of \citet{Perrin-2004oct}. The
      horizontal axes are the spatial frequencies \blambda{}, marked
      in meter per micron, the vertical axes are the linear
      visibilities.}
    \label{fig:vis_curves}
\end{figure*}

Observations of calibrator stars with smaller angular diameters were
interleaved with observations of \TLep{}. We were able to reach an
average observation frequency of about 25min per calibrated point.
This appeared to be crucial to properly
sample the instrumental and atmospheric transfer function (TF). This
also allowed us to manually discard the obvious outlier measurements
without degrading the spatial coverage of our
dataset too much. Additionally, we discarded all observations where the FINITO
locking ratio was lower than 60\%.
Raw visibility and closure phase values were computed using the latest
public version of the \texttt{amdlib} package \citep[version
  2.2,][]{Tatulli-2007mar} and the \texttt{yorick} interface provided
by the Jean-Marie Mariotti Center.  We used our own software to
calibrate the instrumental and atmospheric TF (see
Fig.~\ref{fig:calibrate} for an example). We estimated the uncertainty
on the TF with the dispersion between the consecutive points obtained
for each observation of a calibration star. This dispersion generally
dominated the other sources of error (mainly the 
diameter uncertainty of calibrators) and was propagated to the uncertainties on the
calibrated visibilities and phases obtained for \TLep{}.
Fig.~\ref{fig:uv_plane} shows the final UV-plane coverage of all
observations that successfully passed all steps of the data reduction
and calibration quality control. The east-west
direction is favored because of the geometry of the telescope
triplets used. Visibility curves are displayed in
Fig.~\ref{fig:vis_curves}.

Looking at the interferometric data, we found that the general picture
of \TLep{} significantly departs from a simple disk or Gaussian. It is
composed of at least two different features: one has a characteristic
size $5-7$mas and the other $11-16$mas. The exact size of these
components and the flux ratio depend on the wavelength. This is
qualitatively in agreement with a complex stellar atmosphere
surrounded by molecular shells with wavelength-dependent
opacity. Additionally, the non-zero closure phase 
(departure from $0-180\deg$ up to $90\deg$ in the H-band and
$30\deg$ in the K-band) indicated the presence of asymmetries
either in the shells or in the photosphere itself. Interestingly, we
were not able to reproduce correctly the interferometric data with
simple geometrical models such as Gaussian + Gaussian, or
limb-darkened disk (LDD) + Gaussian, even when disregarding the
closure phases
This was our main motivation to attempt to reconstruct an image.

\section{Image reconstruction}
\label{sec:image}

\begin{figure*}
    \centering
    \includegraphics[scale=0.7]{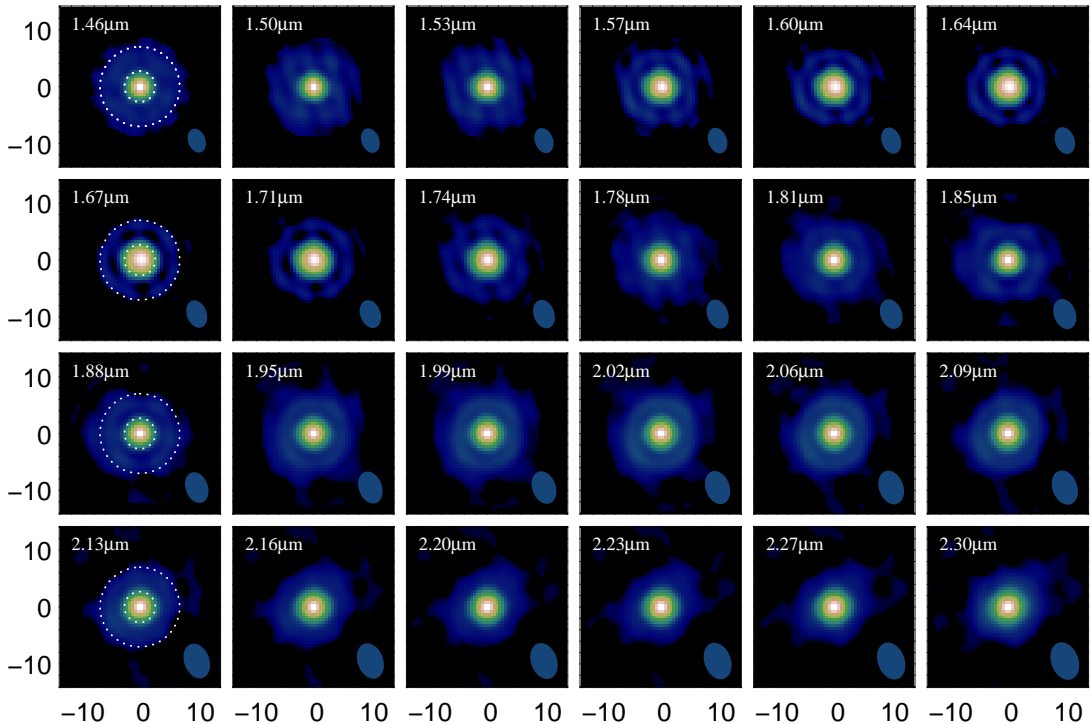}
    \caption{Reconstructed images of \TLep{} with the \mira{} software
      for several AMBER spectral bins across the H and K bands. 
      The interferometric beam size is displayed in the bottom-right part of each image.
      Spatial scale is in mas. The white circles in the first column
      represent the average radius for the molecular layer
      ($\Dlayer\sim15$mas) and for the central star
      ($\Dlayer\sim5.8$mas), extracted from the modeling of
      Sec.~\ref{sec:profile}. It corresponds to the respective diameters
      of 2.5 and 1\,AU \citep[assuming $5.95\pm0.70\,$mas parallax
        from][]{Leeuwen-2007}. The mean surface brightness ratio between
      the photosphere and the molecular environment is around 10\%.
      }
    \label{fig:images}
\end{figure*}

Our dataset is perfectly suited to spectro-spatial inversion, where
the spectral \emph{and} the spatial dimensions are treated together
assuming a certain number of degrees of freedom in each
space. Unfortunately the necessary algorithms are not yet available, and we had
to solve the problem independently for each spectral bin. We performed
the image reconstruction with the \mira{} software
\citep{Thiebaut-2008spie,Cotton-2008spie}. 
In spite of the apparent completeness of the dataset, we expected that it would not be
straightforward to produce images: (i) the geometry of the VLTI array makes the east-west
direction privileged, (ii) the necessary use of closure phases makes
the phase information three times less rich than that of the
visibilities. Therefore, a 2-step image reconstruction strategy was
specifically elaborated. The first step consisted of building a
radially symmetrical image from the data.
Only the baseline length was considered. The amplitude
and sign of these pseudo data (which are all real due to the assumed
symmetry) were derived from the measured power spectrum and sign of
the closure phases respectively. We then used MIRA with a strong
smoothness regularization to reconstruct the first stage images.
The second step uses these brightness distributions
as a support for a quadratic regularization of a 2D brightness map
\citep{LeBesnerais-2008ieee}.  Each of the spectral channels
was processed separately. Resulting final images are
presented in Fig.~\ref{fig:images}.

The recontructed images clearly highlight two components. The brighter
central component corresponds to the 
photosphere, with a linear radius of 100~$R_\odot$ \citep[assuming a
parallax of $5.95\,$mas from][]{Leeuwen-2007}. 
The second component is the molecular layer (molsphere),
characterized by the high spectral dependency of its
apparent morphology. Within the molecular absorption bands
($1.5-1.6\,\microns$, $1.75-2.15\,\microns$, $2.3-2.5\,\microns$), the shell is optically thick
and appears, projected toward the observer, as a circular disk.
Between these bands, absorption by molecules is not as strong,
and the shell becomes optically thin. It appears
as a projected ring encircling the photosphere ($1.6-1.75\,\microns$).
The lack of data points in the second lobe at $2.2\,\microns$ (see Fig.~\ref{fig:vis_curves})
prevented the algorithm from reconstructing the inner gap between the molecular shell and the
photosphere. Instead, it resulted in a decreased contrast between the
molsphere and the photosphere.

Each spectral bin has been reduced and imaged
independently. These images are in good agreement.
It provides a strong indication that the
MIRA algorithm is not inverting the statistical noise present in the
data. Possible correlations may still originate from the calibration
noise (atmospheric turbulence) and the use of common calibrator
stars. However, these sources of errors are unlikely to give rise to
such a coherent pattern as observed in Fig.~\ref{fig:images}. These
images are a strong argument for a shell-like geometry of the
molecules within the atmosphere.

\section{Intensity profile modeling}
\label{sec:profile}

\begin{figure}
    \centering \vspace{-0.3cm}
    \includegraphics[scale=0.6]{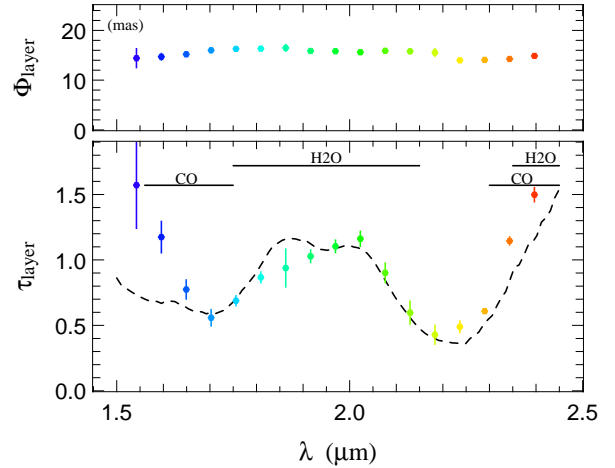}
    \caption{Size and optical depth of the layer surrounding the
      photosphere, as extracted by a fit of the visibility curves
      (Sec.~\ref{sec:profile}). The dashed curve is a 
        fit with H$_2$O and CO opacities assuming
        $\Tlayer=1800\,$K
        (T.~Verhoelst, private communication). A continuum
        opacity of $\tau\approx0.2$ was added.}
    \label{fig:model_perrin}
\end{figure}

In addition to reconstructed images of the molecular shells, we used a
simple radiative transfer model developed by \citet{Perrin-2004oct} to
derive the physical parameters of the system. This model consists of a
uniform disk star that emits radiation with a black body spectrum
($\Dstar,\Tstar$), surrounded by a spherical layer with no geometrical
thickness. The layer is characterized by a size, a temperature and an
optical depth ($\Dlayer,\Tlayer,\Olayer$). Such an analytic model can be
compared to our wealth of observations without requiring time-consuming
computations. We have also preferred to use an ad-hoc
geometrical model rather than a self-consistent hydrodynamical model
in order to easily give some geometrical flexibility to the fitting
process. Other types of models will be discussed in further
publications. We allowed the layer size and optical depth to be
wavelength dependent, but not its temperature nor the stellar
parameters. We were able to reproduce the visibility curves with the
following parameters $\Tstar=3500\,$K, $\Tlayer=1800\,$K,
$\Dstar=5.8\,$mas and the chromatic quantities of
Fig.~\ref{fig:model_perrin}.
According to the good overlay of the visibility curves in
Fig.~\ref{fig:vis_curves}, all spectral features observed in
Fig.~\ref{fig:model_perrin} are real. However, our fit shows important correlations between
$\Tstar{}$, $\Tlayer{}$ and the average level of $\Olayer{}$
($\approx$$1$). Additional spectroscopic inputs may be needed to provide
definite measurements of these quantities.

The model is able to reproduce the wavelength dependency of the
visibility break around $\blambda=20$. Our data strongly suggest that
the star+layer geometry is a good proxy for the real intensity profile
and strengthens the result we had obtained with our imaging
technique.  It also gives a strong argument suggesting that the opacity effect
explains most of the apparent size changes, without requiring
important changes in the layer size (which was wavelength-independent
in previous studies).

Following computations of \citet{Verhoelst-2006feb},
we modeled the spectral dependency of the optical depth as a sum of water vapor 
(mainly contributing in the range $1.75-2.15\,\microns$), and CO 
(below $1.75\,\microns$ and above $2.3\,\microns$).
However, we had to add a continuum opacity of constant value $\approx$$0.2$ to
be able to fairly reproduce the observed curve.
Accordingly, the optical depth of the layer is always greater than
$0.4$ over the complete
spectral range : it is impossible to see the photosphere alone.
The presence of this continuum opacity, as well as the discrepencies in the fit,
will require adapted modeling before being understood.

The simple star+layer  model was able to reproduce almost all the
visibility points with $V>15$\% but clearly failed at higher spatial
frequency. An important feature in the visibility curves is the
increase of contrast with baseline observed around $\blambda=45$ for
the spectral bins around $1.8\,\microns$. It could be tempting to
identify it as a ``visibility second lobe'' following a ``visibility
null'' (confirmed by the closure phases). This visibility lobe is
apparently no longer observed around $1.9\,\microns$ and above. The
only way to continuously remove a visibility null is i) to fill it
with unresolved flux or ii) to displace it to infinite spatial
frequencies by smoothing the corresponding edge, and iii) to add
important asymmetries so that the visibility becomes a complex number
around the null. The two first cases are not plausible in our case
because the visibility around $\blambda=45$ never reaches 0 in the
entire dataset. We conclude that the visibility profiles cannot be
explained by an axisymmetric distribution continuous in
wavelength. Undoubtedly, such an effect would have been missed with a
less complete spectral and/or spatial coverage.

\section{Conclusions}
We believe this letter presents the first interferometric observations
of a Mira star with a dataset almost continuous in the spectral
and in the spatial domain.
Obtaining it in a short amount of time has been made
possible by the use of the AMBER spectrograph associated with the
relocatable ATs of VLTI.

Individual spectral channel image reconstruction was achieved using the
MIRA software. It provides confirmation of the existence of a
spherical molecular shell at $\approx 1$\,AU of the photosphere.
Between $1.5-2.4\,\microns$ we did not find
any spectral region where the brightness profile of \TLep{} can be
explained by a single Gaussian or a single LDD.
Conversely, when using a simple radiative
transfer model we never found an optical
depth less than $0.4$ for the layer surrounding the photosphere.
Rough modeling points to a continuum emission within the molecular
shell, in addition to the CO and H$_2$O features.

\begin{acknowledgements} 
JBLB thanks the complete VLTI team.  We acknowledge the use of the
AAVSO International Database, of the Smithsonian/NASA Astrophysics
Data System, and of the Centre de Donnees Astronomiques de
Strasbourg. All graphics were drawn with the freeware
\texttt{yorick}.
\end{acknowledgements}


\begin{thebibliography}{25}
\expandafter\ifx\csname natexlab\endcsname\relax\def\natexlab#1{#1}\fi

\bibitem[{Lee(2007)}]{Leeuwen-2007}
 2007, Hipparcos, the new reduction of the raw data (Springer)

\bibitem[{{Cotton} {et~al.}(2008){Cotton}, {Monnier}, {Baron}, {Hofmann},
  {Kraus}, {Weigelt}, {Rengaswamy}, {Thi{\'e}baut}, {Lawson}, {Jaffe},
  {Hummel}, {Pauls}, {Schmitt}, {Tuthill}, \& {Young}}]{Cotton-2008spie}
{Cotton}, W., {Monnier}, J., {Baron}, F., {et~al.} 2008, in SPIE Conf.,
  Vol. 7013

\bibitem[{{Haguenauer} {et~al.}(2008){Haguenauer}, {Abuter}, {Alonso},
  {Argomedo}, {Bauvir}, {Blanchard}, {Bonnet}, {Brillant}, {Cantzler}, {Derie},
  {Delplancke}, {Di Lieto}, {Dupuy}, {Durand}, {Gitton}, {Gilli}, {Glindemann},
  {Guniat}, {Guisard}, {Haddad}, {Hudepohl}, {Hummel}, {Jesuran}, {Kaufer},
  {Koehler}, {Le Bouquin}, {L{\'e}v''que}, {Lidman}, {Mardones}, {M{\'e}nardi},
  {Morel}, {Percheron}, {Petr-Gotzens}, {Phan Duc}, {Puech}, {Ramirez},
  {Rantakyr{\"o}}, {Richichi}, {Rivinius}, {Sahlmann}, {Sandrock},
  {Sch{\"o}ller}, {Schuhler}, {Somboli}, {Stefl}, {Tapia}, {Van Belle},
  {Wallander}, {Wehner}, \& {Wittkowski}}]{Haguenauer-2008spie}
{Haguenauer}, P., {Abuter}, R., {Alonso}, J., {et~al.} 2008, in SPIE Conf., 
Vol. 7013

\bibitem[{{Ireland} \& {Scholz}(2006)}]{Ireland-2006apr}
{Ireland}, M.~J. \& {Scholz}, M. 2006, \mnras, 367, 1585

\bibitem[{{Le Besnerais} {et~al.}(2008){Le Besnerais}, {Lacour}, {Mugnier},
  {Thi{\'e}baut}, {Perrin}, \& {Meimon}}]{LeBesnerais-2008ieee}
{Le Besnerais}, G., {Lacour}, S., {Mugnier}, L.~M., {et~al.} 2008, in IEEE
  Journal of Selected Topics in Signal Processing

\bibitem[{{Le Bouquin} {et~al.}(2008){Le Bouquin}, {Abuter}, {Bauvir},
  {Bonnet}, {Haguenauer}, {di Lieto}, {Menardi}, {Morel}, {Rantakyr{\"o}},
  {Schoeller}, {Wallander}, \& {Wehner}}]{LeBouquin-2008spie_b}
{Le Bouquin}, J.-B., {Abuter}, R., {Bauvir}, B., {et~al.} 2008, in SPIE Conf.,
  Vol. 7013

\bibitem[{{Loup} {et~al.}(1993){Loup}, {Forveille}, {Omont}, \&
  {Paul}}]{Loup-1993jun}
{Loup}, C., {Forveille}, T., {Omont}, A., \& {Paul}, J.~F. 1993, \aaps, 99, 291

\bibitem[{{Mennesson} {et~al.}(2002){Mennesson}, {Perrin}, {Chagnon}, {du
  Coud{\'e} Foresto}, {Ridgway}, {Merand}, {Salome}, {Borde}, {Cotton},
  {Morel}, {Kervella}, {Traub}, \& {Lacasse}}]{Mennesson-2002nov}
{Mennesson}, B., {Perrin}, G., {Chagnon}, G., {et~al.} 2002, \apj, 579, 446

\bibitem[{{Millan-Gabet} {et~al.}(2005){Millan-Gabet}, {Pedretti}, {Monnier},
  {Schloerb}, {Traub}, {Carleton}, {Lacasse}, \&
  {Segransan}}]{Millan-Gabet-2005feb}
{Millan-Gabet}, R., {Pedretti}, E., {Monnier}, J.~D., {et~al.} 2005, \apj, 620,
  961

\bibitem[{{Ohnaka}(2004)}]{Ohnaka-2004sep}
{Ohnaka}, K. 2004, \aap, 424, 1011

\bibitem[{{Ohnaka} {et~al.}(2006){Ohnaka}, {Scholz}, \&
  {Wood}}]{Ohnaka-2006feb}
{Ohnaka}, K., {Scholz}, M., \& {Wood}, P.~R. 2006, \aap, 446, 1119

\bibitem[{{Perrin} {et~al.}(1999){Perrin}, {Coud{\'e} du Foresto}, {Ridgway},
  {Mennesson}, {Ruilier}, {Mariotti}, {Traub}, \& {Lacasse}}]{Perrin-1999may}
{Perrin}, G., {Coud{\'e} du Foresto}, V., {Ridgway}, S.~T., {et~al.} 1999,
  \aap, 345, 221

\bibitem[{{Perrin} {et~al.}(2004){Perrin}, {Ridgway}, {Mennesson}, {Cotton},
  {Woillez}, {Verhoelst}, {Schuller}, {Coud{\'e} du Foresto}, {Traub},
  {Millan-Gabet}, \& {Lacasse}}]{Perrin-2004oct}
{Perrin}, G., {Ridgway}, S.~T., {Mennesson}, B., {et~al.} 2004, \aap, 426, 279

\bibitem[{{Petrov} {et~al.}(2007){Petrov}, {Malbet}, {Weigelt}, {Antonelli},
  {Beckmann}, {Bresson}, {Chelli}, {Dugu{\'e}}, {Duvert}, {Gennari},
  {Gl{\"u}ck}, {Kern}, {Lagarde}, {Le Coarer}, {Lisi}, {Millour}, {Perraut},
  {Puget}, {Rantakyr{\"o}}, {Robbe-Dubois}, {Roussel}, {Salinari}, {Tatulli},
  {Zins}, {Accardo}, {Acke}, {Agabi}, {Altariba}, {Arezki}, {Aristidi},
  {Baffa}, {Behrend}, {Bl{\"o}cker}, {Bonhomme}, {Busoni}, {Cassaing},
  {Clausse}, {Colin}, {Connot}, {Delboulb{\'e}}, {Domiciano de Souza},
  {Driebe}, {Feautrier}, {Ferruzzi}, {Forveille}, {Fossat}, {Foy},
  {Fraix-Burnet}, {Gallardo}, {Giani}, {Gil}, {Glentzlin}, {Heiden},
  {Heininger}, {Hernandez Utrera}, {Hofmann}, {Kamm}, {Kiekebusch}, {Kraus},
  {Le Contel}, {Le Contel}, {Lesourd}, {Lopez}, {Lopez}, {Magnard}, {Marconi},
  {Mars}, {Martinot-Lagarde}, {Mathias}, {M{\`e}ge}, {Monin}, {Mouillet},
  {Mourard}, {Nussbaum}, {Ohnaka}, {Pacheco}, {Perrier}, {Rabbia}, {Rebattu},
  {Reynaud}, {Richichi}, {Robini}, {Sacchettini}, {Schertl}, {Sch{\"o}ller},
  {Solscheid}, {Spang}, {Stee}, {Stefanini}, {Tallon}, {Tallon-Bosc}, {Tasso},
  {Testi}, {Vakili}, {von der L{\"u}he}, {Valtier}, {Vannier}, \&
  {Ventura}}]{Petrov-2007mar}
{Petrov}, R.~G., {Malbet}, F., {Weigelt}, G., {et~al.} 2007, \aap, 464, 1

\bibitem[{{Ragland} {et~al.}(2006){Ragland}, {Traub}, {Berger}, {Danchi},
  {Monnier}, {Willson}, {Carleton}, {Lacasse}, {Millan-Gabet}, {Pedretti},
  {Schloerb}, {Cotton}, {Townes}, {Brewer}, {Haguenauer}, {Kern}, {Labeye},
  {Malbet}, {Malin}, {Pearlman}, {Perraut}, {Souccar}, \&
  {Wallace}}]{Ragland-2006nov}
{Ragland}, S., {Traub}, W.~A., {Berger}, J.-P., {et~al.} 2006, \apj, 652, 650

\bibitem[{{Samus} {et~al.}(2004){Samus}, {Durlevich}, \& {et
  al.}}]{Samus-2004nov}
{Samus}, N., {Durlevich}, O.~V., \& {et al.} 2004, VizieR Online Data Catalog,
  2250, 0

\bibitem[{{Tatulli} \& {AMBER consortium}(2007)}]{Tatulli-2007mar}
{Tatulli}, E. \& {AMBER consortium}. 2007, \aap, 464, 29

\bibitem[{{Thiebaut}(2008)}]{Thiebaut-2008spie}
{Thiebaut}, E. 2008, in SPIE Conf., Vol. 7013

\bibitem[{{Thompson} {et~al.}(2002){Thompson}, {Creech-Eakman}, \& {van
  Belle}}]{Thompson-2002sep}
{Thompson}, R., {Creech-Eakman}, M.~J., \& {van Belle}, G. 2002, \apj, 577, 447

\bibitem[{{Verhoelst} {et~al.}(2006){Verhoelst}, {Decin}, {van Malderen},
  {Hony}, {Cami}, {Eriksson}, {Perrin}, {Deroo}, {Vandenbussche}, \&
  {Waters}}]{Verhoelst-2006feb}
{Verhoelst}, T., {Decin}, L., {van Malderen}, R., {et~al.} 2006, \aap, 447, 311

\bibitem[{{Whitelock} {et~al.}(2000){Whitelock}, {Marang}, \&
  {Feast}}]{Whitelock-2000dec}
{Whitelock}, P., {Marang}, F., \& {Feast}, M. 2000, \mnras, 319, 728

\bibitem[{{Wittkowski} {et~al.}(2008){Wittkowski}, {Boboltz}, {Driebe}, {Le
  Bouquin}, {Millour}, {Ohnaka}, \& {Scholz}}]{Wittkowski-2008feb}
{Wittkowski}, M., {Boboltz}, D.~A., {Driebe}, T., {et~al.} 2008, \aap, 479, L21

\bibitem[{{Wittkowski} {et~al.}(2007){Wittkowski}, {Boboltz}, {Ohnaka},
  {Driebe}, \& {Scholz}}]{Wittkowski-2007jul}
{Wittkowski}, M., {Boboltz}, D.~A., {Ohnaka}, K., {et~al.}
  2007, \aap, 470, 191

\bibitem[{{Woodruff} {et~al.}(2004){Woodruff}, {Eberhardt}, {Driebe},
  {Hofmann}, {Ohnaka}, {Richichi}, {Schertl}, {Sch{\"o}ller}, {Scholz},
  {Weigelt}, {Wittkowski}, \& {Wood}}]{Woodruff-2004jul}
{Woodruff}, H.~C., {Eberhardt}, M., {Driebe}, T., {et~al.} 2004, \aap, 421, 703

\bibitem[{{Woodruff} {et~al.}(2008){Woodruff}, {Tuthill}, {Monnier}, {Ireland},
  {Bedding}, {Lacour}, {Danchi}, \& {Scholz}}]{Woodruff-2008jan}
{Woodruff}, H.~C., {Tuthill}, P.~G., {Monnier}, J.~D., {et~al.} 2008, \apj,
  673, 418

\end{thebibliography}


\end{document}